\documentclass[12pt]{article}
\usepackage{epsf,a4wide}

\begin{document}
\begin{center}
{\Large{\bf  Spin effects in diffractive $ Q \bar Q$ production at
BNL eRHIC}} \vskip 2mm S.V.~Goloskokov
 \vskip 2mm {\small {
\it Bogoliubov Laboratory of Theoretical Physics, Joint Institute
for Nuclear Research,\\ Dubna 141980, Moscow region, Russia }
\\
 {\it E-mail: goloskkv@thsun1.jinr.ru }}
\end{center}

\begin{abstract}
We discuss  quark-antiquark leptoproduction  within a QCD
two-gluon exchange model at small $x$.  The double spin
asymmetries for longitudinally polarized leptons and transversely
polarized protons in diffractive  $Q \bar Q$ production are
analysed at eRHIC energies. The predicted $A_{lT}$ asymmetry  is
large and can be used to obtain information on the polarized
generalized gluon distributions in the proton.
\end{abstract}

PACS: 12.38.Bx, 13.87.-a, 13.88.+e

\section{Introduction}
Investigation of  hadron photo and leptoproduction at high
energies and small Bjorken $x$ is a problem of considerable
interest. In this region the predominant contribution is
determined by Pomeron exchange. Intensive experimental study of
diffractive processes was performed in DESY (see, e.g.,
\cite{zeus97,jpsi1,h1_99,dijet} and references therein). They give
 reach information on the Pomeron structure and properties of
spin-average gluon distributions in the proton. The spin structure
of the Pomeron was analysed by different authors (see, e.g.,
\cite{gol_mod,gol_kr,butt} and reference therein). Such spin
effects which do not vanish at high energies might be determined
by the large distance contributions in the hadron structure which
lead to a complicated spin-dependent form of the Pomeron--proton
vertex. The gluon-loop corrections to the Pomeron coupling with
the hadron might produce  other sources of the spin-flip part in
this coupling \cite{gol_glu}. Manifestation of spin-dependent
Pomeron can be investigated in the elastic pp scattering at low
$t$ \cite{kopel}, near diffraction minimum \cite{gol02} and in
polarized diffractive hadron leptoproduction \cite{golostr}.

Spin effects in  diffractive processes at small $x$ should be
studied at accelerators with polarized beams.   The Pomeron in QCD
is related with a two-qluon colorless exchange \cite{low}, and in
polarized experiments the spin-dependent two-gluon coupling with
the proton can be analyzed. In diffractive hadron photoproduction
the momentum carried by the two-gluon system is usually not equal
to zero. In this case, the two-gluon coupling with the proton
might be expressed in terms of the generalized parton distribution
(GPD) in the nucleon \cite{rad-j}.

The spin-dependent gluon distributions contain significant
information on the spin structure of the proton.  The important
role in investigation of polarized gluon GPD should play the
diffractive $Q \bar Q$ leptoproduction at small $x$. Information
on this process can be obtained from dijet events in lepton-proton
interaction. Theoretical analysis of these reactions was carried
out in \cite{die95, bart96, ryskin97, schaef}. It was shown that
the cross sections of diffractive quark- antiquark production are
expressed in terms of the same gluon distributions as in the case
of  vector meson production. Thus, the diffractive $Q \bar Q$
leptoproduction at small $x$ really might be an excellent tool
that can be used to study the gluon GPD  at small $x$. Spin
effects in $Q \bar Q$ production in lepton-proton reaction at
small $x<0.1$, where the gluon contribution prevails, was studied,
for example in \cite{golostr}.  The double spin $A_{lT}$
asymmetries in $Q \bar Q$ production for a longitudinally
polarized lepton and a transversely polarized proton were analysed
within the two- gluon exchange model. The predicted $A_{lT}$
asymmetry is  not small and is sensitive to the spin-dependent
part of the two-gluon coupling with the proton.

In future one of the best places to study spin effects in
diffractive lepton-proton reactions will be the eRHIC accelerator
\cite{abhay} with a polarized lepton beam which will collide with
the RHIC beam. The  eRHIC kinematics, where the photon momentum is
much smaller than the proton momentum, is asymmetric. Such a
kinematics is similar to the HERA accelerator and is suitable to
study hard diffractive events. The polarized protons from RHIC can
be used to study spin effects in diffractive hadron
leptoproduction at small $x$. In this paper, we discuss a
possibility to study $A_{lT}$ asymmetry in diffractive $Q \bar Q$
production in future experiments at eRHIC in order to receive
information on the polarized gluon distributions (preliminary
results can be found in \cite{prag03}). In the second section, we
analyse the kinematics of the final particle in this reaction. It
is shown that  the final proton moves practically in the same
direction as the initial one. Jets from the final quarks have
large angles and should be analysed by the eRHIC detector. In the
third section, the main theoretical equations which determine the
spin asymmetry in $Q \bar Q$ production are presented. Predictions
for spin asymmetries at eRHIC energies are made in  section 4. The
expected asymmetries are not small,  about a few per cent. This
shows a possibility to study $A_{lT}$ asymmetry at the future
eRHIC experiments where the information on the polarized gluon GPD
can be obtained.

\section{Kinematics of diffractive $Q \bar Q$ leptoproduction at\\ eRHIC}

Let us study the diffractive quark initiated dijet production in
lepton-proton reactions
\begin{equation}
\label{react} l+p \to l+p +H
\end{equation}
at high energies in a lepton-proton system. The hadronic state $H$
contains  two diffractively produced quarks which are observed as
two final jets.
\bigskip
\begin{figure}[h]
\centering \mbox{\epsfysize=40mm\epsffile{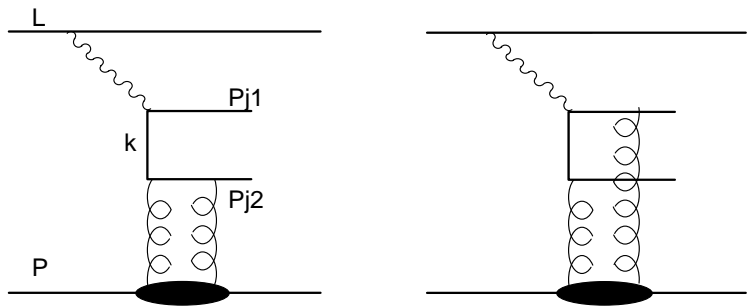}}
\caption{Two-gluon contribution to $Q \bar Q$ production}
\label{Fgg}
\end{figure}

The graphs with $t$-channel gluon exchange in diffractive $Q \bar
Q$ leptoproduction are shown in Fig. 1. This contribution
predominates at small $x \leq 0.1$ for light quark  and essential
for heavy (charm) quark production, because the charm component in
the proton is small. The reaction (\ref{react}) is described in
terms of the kinematic variables which are defined by
\begin{eqnarray}
\label{momen} q^2= (L-L')^2=-Q^2,\;t=r_P^2=(P-P')^2, \nonumber
\\  y=\frac{P \cdot q}{L  \cdot P},\;x=\frac{Q^2}{2P  \cdot q},\;
x_P=\frac{q \cdot (P-P')}{q \cdot P},\; \beta=\frac{x}{x_P},
\end{eqnarray}
where $L, L'$ and $P, P'$ are the initial and final lepton and
proton momenta, respectively, $Q^2$ is the photon virtuality,
$r_P$ is the momentum carried by the Pomeron, $t$ is a momentum
transfer squared and $x$ is a Bjorken variable. In (\ref{momen})
$y$ and $x_P$ represent the fractions of the longitudinal momenta
of the lepton and proton carried by the photon and Pomeron,
respectively. The energy of the lepton--proton system  reads as $
s = (L+P)^2$. The effective mass of a produced quark system is
equal to $M_X^2=(q+r_P)^2\sim x_P y s -Q^2$ and can be large. The
variable $\beta$ is equal to
\begin{equation}
\beta=x/x_P \sim Q^2/(M_X^2+Q^2)
\end{equation}
and can vary from 0 to 1.

We use here the light--cone variables that are determined as
$a_\pm=a_0 \pm a_z$. In calculation, the center of mass system is
used, where the initial lepton and proton momenta are going along
the $z$ axis. They have the form
\begin{equation}
\label{lp} L=(p_{+},\frac{\mu^2}{p_{+}},\vec 0),\quad
P=(\frac{m^2}{p_{+}},p_{+},\vec 0).
\end{equation}
Here $\mu$ and $m$ are the lepton and proton mass. The momenta of
the photon and the Pomeron can be written as follows:
\begin{eqnarray}\label{kinem}
q&=&(y p_{+},-\frac{Q^2}{p_{+}},\vec q_\bot),\quad
|q_\bot|=\sqrt{Q^2 (1-y)};
\nonumber \\
r_P&=&(-\frac{|t|}{p_{+}},x_P p_{+},\vec r_\bot),\;\;
|r_\bot|=\sqrt{|t| (1-x_P)}.
\end{eqnarray}

The final quark momenta $p_1$, $p_2$ and the momentum of the
off-mass-shell quark $k$ (see Fig.1) are determined from the
mass-shell conditions $p_1^2=m_q^2$, $p_2^2=m_q^2$. The vector $k$
is mainly transverse: $k^2 \sim -k_\perp^2$. The momenta of the
observed jets which are equal to the quark momenta have the
following form: \cite{golostr}
\begin{eqnarray}\label{p1p2}
p_{J1}& \sim &\left(y p_{+} -\frac{|t|}{p_{+}}- \frac{m_q^2+(\vec
r_\perp+\vec k_\perp)^2}{p_{+} x_P}, \frac{m_q^2+(\vec
q_\perp-\vec k_\perp)^2}{p_{+} y}, (\vec q_\perp-\vec k_\perp)
\right),
\nonumber \\
p_{J2}& \sim &\left(\frac{m_q^2+(\vec r_\perp+\vec
k_\perp)^2}{p_{+} x_P}, x_P
p_{+}-\frac{Q^2}{p_{+}}-\frac{m_q^2+(\vec q_\perp-\vec
k_\perp)^2}{p_{+} y}, (\vec r_\perp+\vec k_\perp) \right).
\end{eqnarray}

\begin{figure}[h]
\centering \mbox{\epsfysize=30mm\epsffile{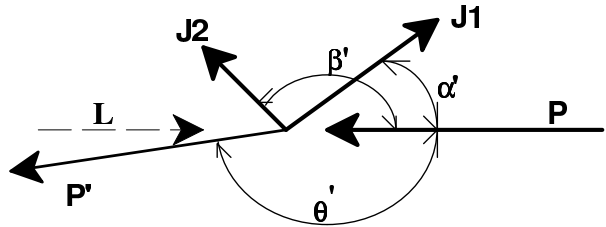}}
\caption{Final hadron kinematics in $Q \bar Q$ production at eRHIC
system} \label{jj}
\end{figure}

We can write the same momenta at the eRHIC asymmetric system, Fig.
\ref{jj}. We use here the standard variables $v=(v_0,v_\perp,v_z)$
and do not consider in details of scattering in the transverse
plane, for simplicity. Only one transverse component  which is
determined by $v_\perp=\sqrt{v_x^2+v_y^2}$ is used in the vectors.
In this approximation, we can write
\begin{eqnarray}\label{esys}
L& = &\left(p_l,0,p_l\right),\nonumber \\
L'& = &\left(p^f_l,p^f_l \sin{\gamma'},p^f_l \cos{\gamma'}\right),\nonumber \\
P& = &\left(\sqrt{m^2+p^2},0,-p\right),\nonumber \\
P'& = &\left(\sqrt{m^2+p_f^2},p_f \sin{\theta'},-p_f \cos{\theta'}\right),\nonumber \\
P_{J1}& = &\left(\sqrt{m_Q^2+J_1^2},J_1 \sin{\alpha'},J_1 \cos{\alpha'}\right),\nonumber \\
P_{J2}& = &\left(\sqrt{m_Q^2+J_2^2},J_2 \sin{\beta'},J_2
\cos{\beta'}\right).
\end{eqnarray}
One can estimate the momenta and  the scattering angles in the
eRHIC system in terms of variables (\ref{momen}) by comparison of
scalar productions of momenta in different systems. For the final
lepton and hadron momenta and there angles we have
\begin{eqnarray}\label{re}
p^f_l\sim 2\, p_l\,\frac{1-y}{1+\cos{\gamma'}}, \; \hspace{1cm}
\cos{\gamma'}\sim
\frac{4\, p_l^2\, (1-y)-Q^2}{4 \,p_l^2\, (1-y)+Q^2};\nonumber\\
 p_f\sim p (1-x_P)-\frac{|t|+2 m^2 x_P}{4 p (1-x_P)}, \;
\cos{\theta'} \sim -1 +\frac{|t|+ m^2 (1+ x_P)}{4 p^2 (1-x_P)}.
\end{eqnarray}
For jet momenta and its angles  the following solutions were
found:
\begin{eqnarray}\label{je}
J_1\sim y p_l+\frac{m_q^2+(q_\perp-k_\perp)^2)}{4 y p_l},\
cos{\alpha'}\sim \frac{(2 y p_l)^2 -m_q^2-(q_\perp-k_\perp)^2}{(2
y p_l)^2+m_q^2+(q_\perp-k_\perp)^2};\nonumber\\
J_2\sim x_P p -\frac{m_q^2+y Q^2 +(q_\perp-k_\perp)^2}{4 y p_l},\;
cos{\beta'}\sim -1+\frac{m_q^2+(k_\perp+r_\perp)^2}{2(x_P p)^2}.
\end{eqnarray}

In the eRHIC  system  the lepton momentum should be
$p_l=5\mbox{GeV}$ or $p_l=10\mbox{GeV}$. For the  proton momentum
two possibilities are analysed $p=100\mbox{GeV}$ and
$p=250\mbox{GeV}$. The energy in the lepton-proton system is $s
\sim 4 p_l p+m^2$ and for our choice the minimum and maximum
energy for eRHIC will be $\sqrt{s} \sim 50 \mbox{GeV}$ and
$\sqrt{s} \sim 100 \mbox{GeV}$. To estimate the kinematics of the
final particles at eRHIC, we use some typical values of the
variables determined in (\ref{momen}): $x_P=0.05, y=0.3,
Q^2=5\mbox{GeV}^2, k_\perp^2=2\mbox{GeV}^2$ and $r_\perp$ is
small.
 In this case, the final lepton momentum and angle can be
estimated for $p_l=5\mbox{GeV}$ as
\begin{equation}\label{l5}
p^f_l\sim 3.75 \mbox{GeV}, \; \gamma' \sim 30^o.
\end{equation}
For $p_l=10\mbox{GeV}$ we find
\begin{equation}\label{l10}
p^f_l\sim 7.13 \mbox{GeV}, \; \gamma' \sim 15^o.
\end{equation}
The kinematics of the final hadron production   for energies
$\sqrt{s} \sim 50 \mbox{GeV}$ and $\sqrt{s} \sim 100 \mbox{GeV}$
is shown in the table. The results are presented for the light
quarks. We consider here two cases when the vector $k_\perp$ is
parallel ($+k$) or antiparallel ($-k$) to the transverse component
of the photon momentum $q_\perp$.
\begin{figure}[h]
\centering \mbox{\epsfysize=60mm\epsffile{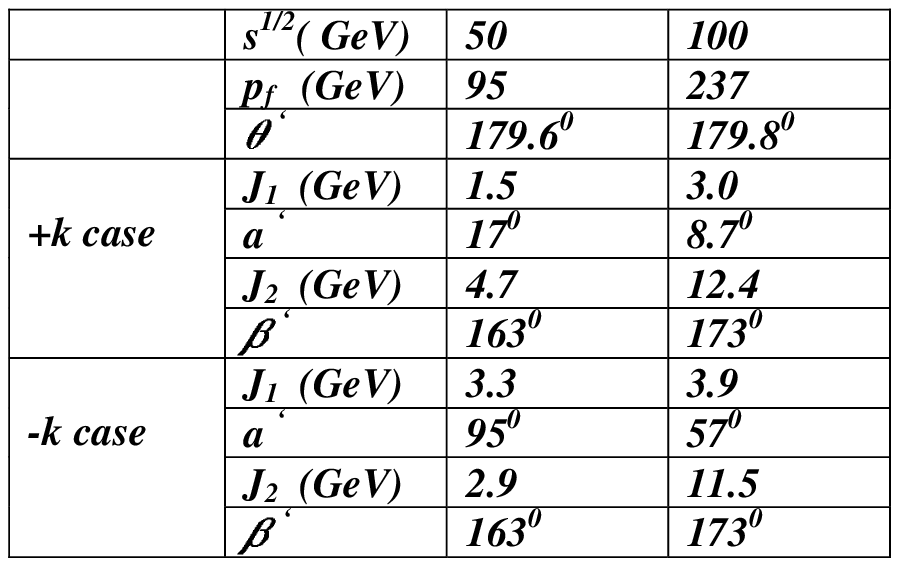}} \label{tab}\\
{Table: Kinematics of the final hadron production. Light quark
case}
\end{figure}

We can see from the Table that the final proton is going inside
the cone with angle $\theta_p^f=(\pi-\theta') < 0.4^o$. It moves
practically in the same direction as the initial proton and it is
possibly difficult to detect it. It is shown in the Table that the
final quarks have large angles $\alpha'\gg \theta', \beta'\gg
\theta'$. Thus the final proton and jet angles are very different
and final jets should be detected by the eRHIC.

\section{Polarized diffractive $Q \bar Q$ leptoproduction}

To study spin effects in diffractive hadron production, one must
know the structure of the two-gluon coupling with the proton at
small $x$. The two-gluon coupling with the proton which describes
transverse spin effects in the proton can be parametrized in the
form \cite{golostr}
\begin{eqnarray}\label{ver}
V_{pgg}^{\alpha\beta}(p,t,x_P,l_\perp)&=& B(t,x_P,l_\perp)
(\gamma^{\alpha} p^{\beta} + \gamma^{\beta} p^{\alpha}) \nonumber\\
&+&\frac{i K(t,x_P,l_\perp)}{2 m} (p^{\alpha} \sigma ^{\beta
\gamma} r_{\gamma} +p^{\beta} \sigma ^{\alpha \gamma}
r_{\gamma})+...  .
\end{eqnarray}
Here $m$ is the proton mass. In the matrix structure (\ref{ver})
we wrote only the terms with the maximal powers of a large proton
momentum $p$ which are symmetric in the gluon indices
$\alpha,\beta$. The structure proportional to $B(t,...)$
determines the spin-non-flip contribution. The term $\propto
K(t,...)$ leads to the transverse spin-flip at the vertex. These
distribution functions should be connected with $F_\zeta(x)$,
$K_\zeta(x)$ which  describe the spin-average and transverse spin
distributions, respectively. If one considers the longitudinal
spin effects, the asymmetric structure $\propto \gamma_\rho
\gamma_5$ should be included in (\ref{ver}).

There are some models (see e.g. \cite{gol_mod,gol_kr,kopel} and
references therein) that provide spin-flip effects which do not
vanish at high energies. The models \cite{gol_mod,gol_kr} describe
the experimental data on single spin transverse asymmetry $A_N$
\cite{krish} quite well. The   spin asymmetries in $pp$ scattering
at RHIC energies (pp2pp experiment) were predicted to be not small
\cite{akch}. The conclusion was made that the weak energy
dependence of spin asymmetries in exclusive reactions is not now
in contradiction with  experiment \cite{gol_mod,akch}. Thus, the
ratio   $|\tilde K|/|\tilde B|$  might have a weak $x$ -dependence
and  be about $0.1$, as was found in \cite{gol_mod,gol_kr}. This
value
 will be used here in  estimations of the
asymmetry in diffractive $Q \bar Q$  production.

The spin-average and spin dependent cross sections with
longitudinal polarization of a lepton and a  transverse proton
polarization are determined by
\begin{equation}
\label{spm} d \sigma(\pm) =\frac{1}{2} \left( d
\sigma({\rightarrow} {\Downarrow) \pm  d \sigma({\rightarrow}
{\Uparrow}})\right).
\end{equation}
To calculate  the cross sections, we  integrate the amplitudes
squared over the $Q \bar Q$ phase space. The  cross section of
diffractive processes are expressed in terms of the soft gluon
coupling (\ref{ver}), which is convoluted with the hard hadron
production amplitude. The spin-average and spin-dependent cross
section can be written in the form
\begin{equation}
\label{sigma} \frac{d^5 \sigma(\pm)}{dQ^2 dy dx_p dt dk_\perp^2}=
\left(^{(2-2 y+y^2)} _{\hspace{3mm}(2-y)}\right)
 \frac{C(x_P,Q^2) \; N(\pm)}
{\sqrt{1-4(k_\perp^2+m_q^2)/M_X^2}}.
\end{equation}
Here $C(x_P,Q^2)$ is a normalization function which is common for
the spin average and spin dependent cross section, and $N(\pm)$ is
determined by a sum of graphs integrated over the gluon momenta.

The $N(+)$  function, which determines the spin-average cross
section, can be written as
\begin{equation}\label{np}
N(+)=\left(|\tilde B|^2+|t|/m^2 |\tilde K|^2 \right)
\Pi^{(+)}(t,k_\perp^2,Q^2).
\end{equation}
The function $\Pi^{(+)}$ has a complicated form and was calculated
numerically. The details of calculations can be found in
\cite{golostr}.

The cross section (\ref{sigma},\ref{np}) is expressed  in terms of
the functions $B$ and $K$  integrated over the gluon momentum
\begin{eqnarray}\label{bqq}
\tilde B \sim \int^{l_\perp^2<k_0^2}_0 \frac{d^2l_\perp
(l_\perp^2+\vec l_\perp \vec r_\perp) }
{(l_\perp^2+\lambda^2)((\vec l_\perp+\vec r_\perp)^2+\lambda^2)}
B(t,l_\perp^2,x_P,...) =  {\cal F}^g_{x_P}(x_P,t,k_0^2)\nonumber\\
\tilde K \sim \int^{l_\perp^2<k_0^2}_0 \frac{d^2l_\perp
(l_\perp^2+\vec l_\perp \vec r_\perp) }
{(l_\perp^2+\lambda^2)((\vec l_\perp+\vec r_\perp)^2+\lambda^2)}
K(t,l_\perp^2,x_P,...) =  {\cal K}^g_{x_P}(x_P,t,k_0^2),
\end{eqnarray}
where $k_0^2 \sim \frac{k_\perp^2+m_q^2}{1-\beta}$
\cite{bart96,ryskin97}. The connection with the gluon GPD
\cite{golostr} is achieved in equations (\ref{bqq}). This shows
that the functions $B (t , l_\perp...)$ and $K (t , l_\perp. . .)$
are the nonintegrated gluon distributions.

The spin-dependent cross section is determined by the interference
between spin-average and spin-dependence distributions. It was
found in \cite{golostr} that the $N(-)$ function in (\ref{sigma})
contains two terms which are proportional to the scalar production
$\vec Q \vec S_\perp$ and to $\vec k_\perp \vec S_\perp$ where
$S_\perp$ is  transverse polarization of the proton
\begin{equation}\label{nm}
N(-)=\sqrt{\frac{|t|}{m^2}} \left(\tilde B \tilde K^*+\tilde B^*
\tilde K\right) [ \frac{(\vec Q \vec S_\perp)}{m}
\Pi^{(-)}_Q(t,k_\perp^2,Q^2)
 +\frac{(\vec k_\perp \vec
S_\perp)}{m} \Pi^{(-)}_k(t,k_\perp^2,Q^2)].
\end{equation}
The contributions of these terms to the asymmetry can be analysed
independently. Really, if one does not consider the azimuthal jets
kinematics and integrate over $k_\perp$, only the term
proportional to $\vec Q \vec S_\perp$ will contribute to the
asymmetry.

\section{Predictions for  $Q \bar Q$ Leptoproduction}
The spin-average cross section of the vector meson production at
small momentum transfer is approximately proportional to the
$|\tilde B|^2$ function (\ref{np}) which is connected with the
generalized gluon distribution ${\cal F}^g$.  We use here the
simple parameterization of the GPD as a product of the
$t$-dependent form factor and the ordinary gluon distribution
\begin{equation}
\label{b_g}
\tilde B(t,x_P, \bar Q^2) =F_B(t) \left( x_P
G(x_P,\bar Q^2)  \right).
\end{equation}
A more general form of GPD (see e.g. \cite{prag03,dis03}) can be
analysed. However, some enhancement factor which appears in this
case will be mainly canceled in the asymmetry. The form factor
$F_B(t)$ in (\ref{b_g}) is chosen as an electromagnetic form
factor of the proton. Such a simple choice can be justified by the
fact that the Pomeron--proton vertex might be similar to the
photon--proton coupling \cite{nach}
\begin{equation}
\label{fp} F_B(t) \sim F^{em}_p(t)=\frac{(4 m_p^2+2.8 |t|)}{(4
m_p^2+|t|)( 1+|t|/0.7GeV^2)^2}.
\end{equation}

 The energy dependence of the cross sections at small $x$ is determined
by the Pomeron contribution to the gluon distribution function
\begin{equation}
\label{g_x} \left(x_P G(x_P,\bar Q^2) \right) \sim
\frac{const}{x_P^{\alpha_p(t)-1}}.
\end{equation}
Here $\alpha_p(t)$ is a Pomeron trajectory which has the form
\begin{equation}\label{pom}
  \alpha_p(t)=1+\epsilon+ \alpha' t
\end{equation}
with $\epsilon=0.17$ and $ \alpha'=0$. These values are in
accordance with the fit of the diffractive $J/\Psi$ production by
ZEUS \cite{zeus97}. For $Q_0^2=\bar Q^2 \sim 4\mbox{GeV}^2$ the
ordinary gluon distribution can be approximated at small $x$ by
\cite{gehr}
\begin{equation}
 \left(x_P G(x_P,Q_0^2) \right) \sim 1.94 x^{-0.17}.
\end{equation}

The $A_{lT}$ asymmetry of hadron production  is determined as a
ratio of spin-dependent and spin-average cross sections
(\ref{sigma},\ref{np},\ref{nm})
\begin{equation}
\label{asylt} A_{lT}=\frac{\sigma(-)}{\sigma(+)}.
\end{equation}
It can be seen that at small momentum transfer the asymmetry is
approximately proportional to the ratio of polarized and
spin--average gluon distribution functions
\begin{equation}\label{cltqq}
 A_{LT}^{Q \bar Q} \sim C^{Q \bar Q} \frac{|\tilde K|}{|\tilde B|}=
 C^{Q \bar Q} \frac{{\cal K}^g_{x_P}(x_P)}
 {{\cal F}^g_{x_P}(x_P)}.
\end{equation}
In estimations of asymmetry we shall use the same ratio of
spin-dependent and spin-average gluon structures, as in the case
of elastic scattering
\begin{equation}\label{ratio}
\frac{|\tilde K|}{|\tilde B|} \sim 0.1
\end{equation}
for simplicity. This means that in this case the coefficient $C^{Q
\bar Q}$ in (\ref{cltqq}) is equal to $10\; A_{LT}^{Q \bar Q}$.
\bigskip
\begin{figure}[h]
\centering \mbox{\epsfysize=80mm\epsffile{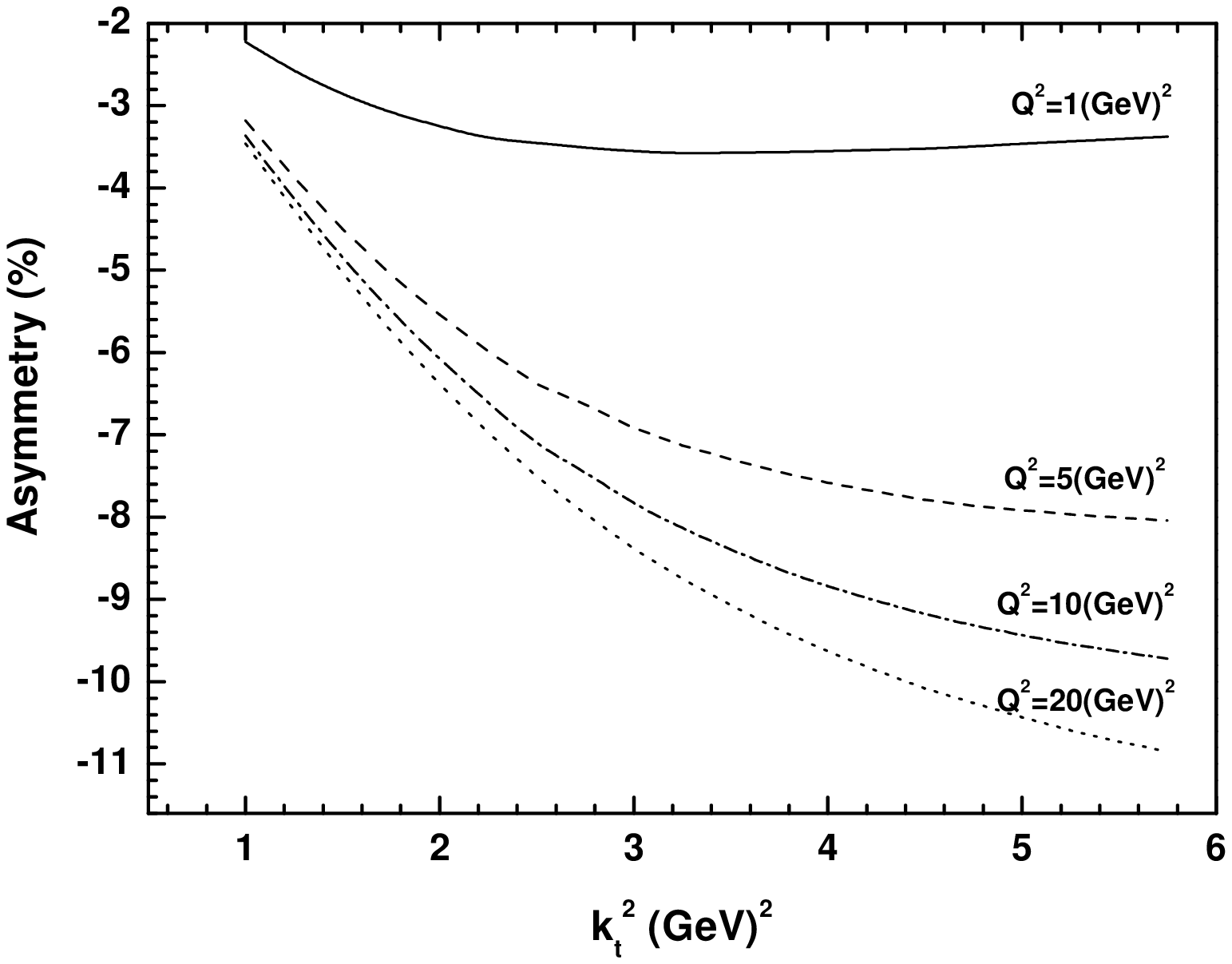}}
\caption{The $A^k_{lT}$ asymmetry in diffractive light $Q \bar Q$
production at $\sqrt{s}=50 \mbox{GeV}$ for $x_P=0.05$, $y=0.3$,
$|t|=0.3 \mbox{GeV}^2$} \label{kt_l}
\end{figure}
\begin{figure}[ht]
\centering \mbox{\epsfysize=80mm\epsffile{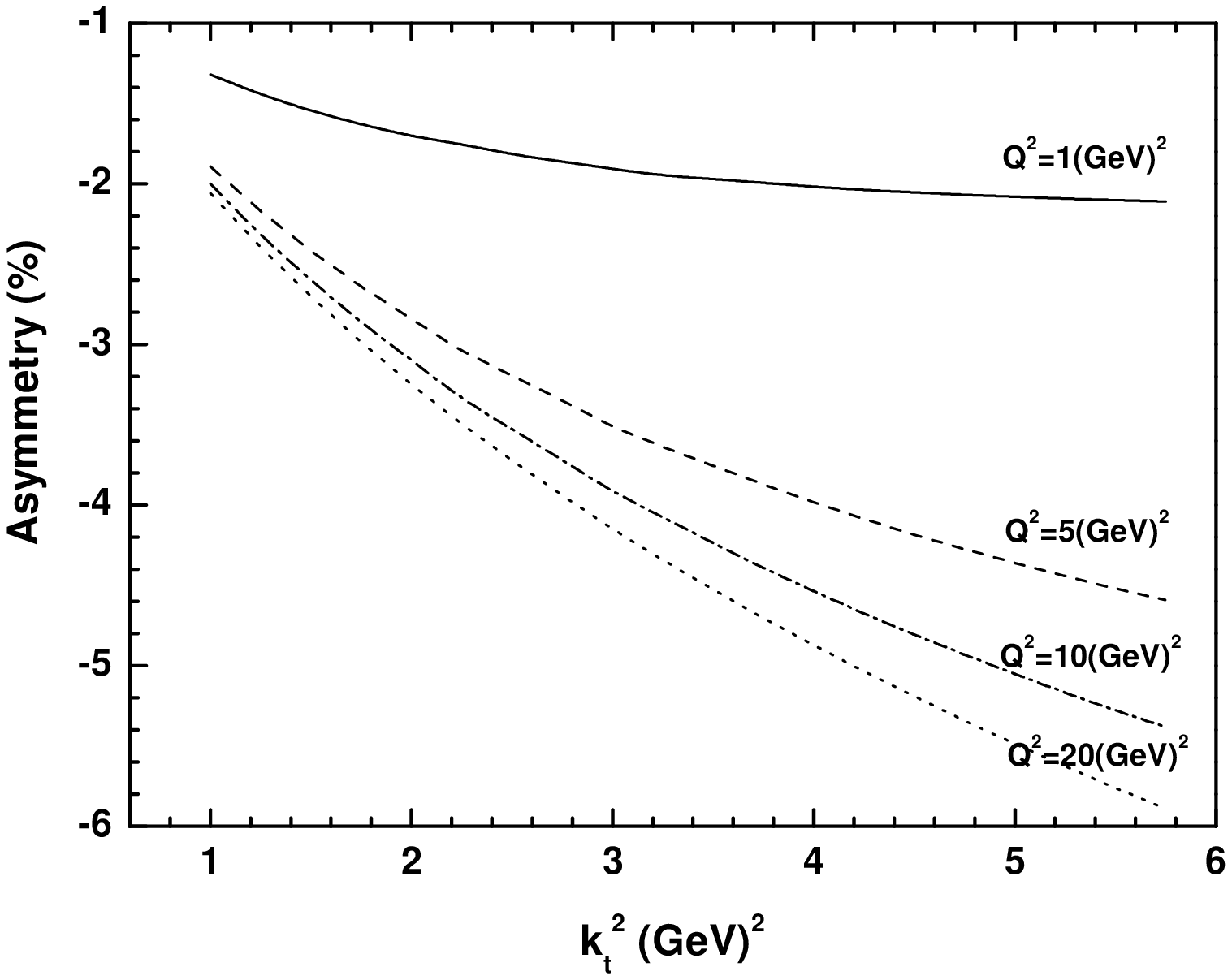}}
\caption{The $A^k_{lT}$ asymmetry in diffractive heavy $Q \bar Q$
production at $\sqrt{s}=50 \mbox{GeV}$ for $x_P=0.05$, $y=0.3$,
$|t|=0.3 \mbox{GeV}^2$} \label{kt_h}
\end{figure}
The term  proportional to $\vec k_\perp \vec S_\perp$ in the
asymmetry ($A^k_{lT}$) will be analyzed for the case when the
transverse jet momentum $\vec k_\perp$ is parallel to the target
polarization $\vec S_\perp$. The asymmetry is maximal in this
case. To observe this contribution to asymmetry, it is necessary
to distinguish experimentally the quark and antiquark jets.  This
can be realized presumably by analyses of  charge of the leading
particles in the jet which should be connected  with the charge of
quark produced in the process. If one does not separate events
with $\vec k_\perp$ which is parallel to $\vec S_\perp$ for the
quark jet, e.g., the resulting asymmetry will be zero because the
transverse momentum of the quark and antiquark are equal and
opposite in sign.

 The predicted asymmetry \cite{prag03} for light quark
production at  energy $\sqrt{s}=50 \mbox{GeV}$ is shown in Fig.
\ref{kt_l}. The asymmetry for heavy $c \bar c$  production is
approximately of the same order of magnitude  (Fig. \ref{kt_h}).
The model used for the ratio of gluon distribution (\ref{ratio})
leads to the weak energy dependence of asymmetry. As a result, the
asymmetry for $\sqrt{s}=100 \mbox{GeV}$ should be practically the
same. The predicted asymmetries are not small, about 5-10 \%. This
shows a possibility of studying the polarized gluon distribution
${\cal K}^g_\zeta(x)$ in a future eRHIC experiment.

\begin{figure}
\centering \mbox{\epsfysize=80mm\epsffile{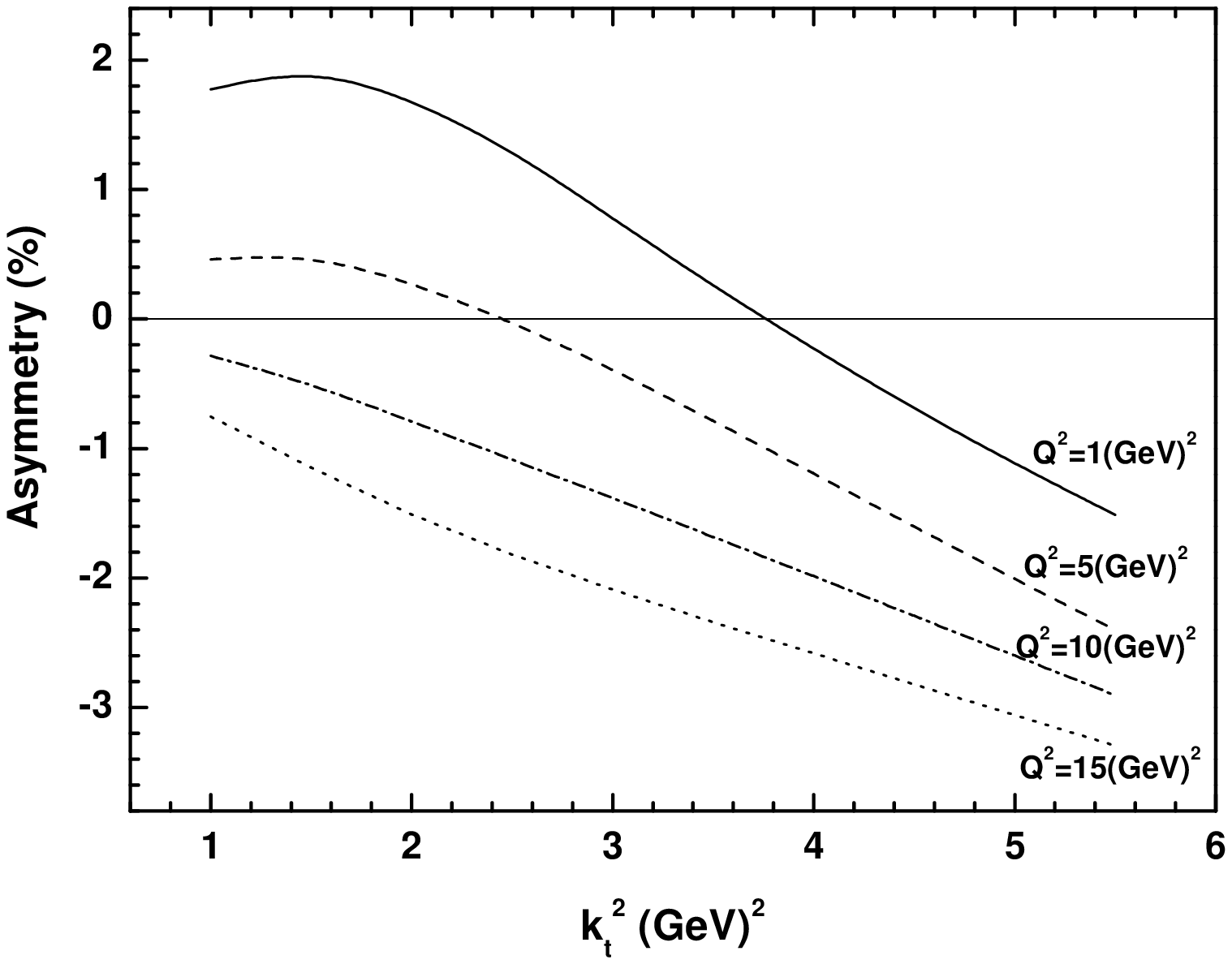}}
\caption{The $A^Q_{lT}$ asymmetry in diffractive light $Q \bar Q$
production at $\sqrt{s}=50 \mbox{GeV}$ for $x_P=0.05$, $y=0.3$,
$|t|=0.3 \mbox{GeV}^2$} \label{qt_l}
\end{figure}

\begin{figure}
\centering \mbox{\epsfysize=80mm\epsffile{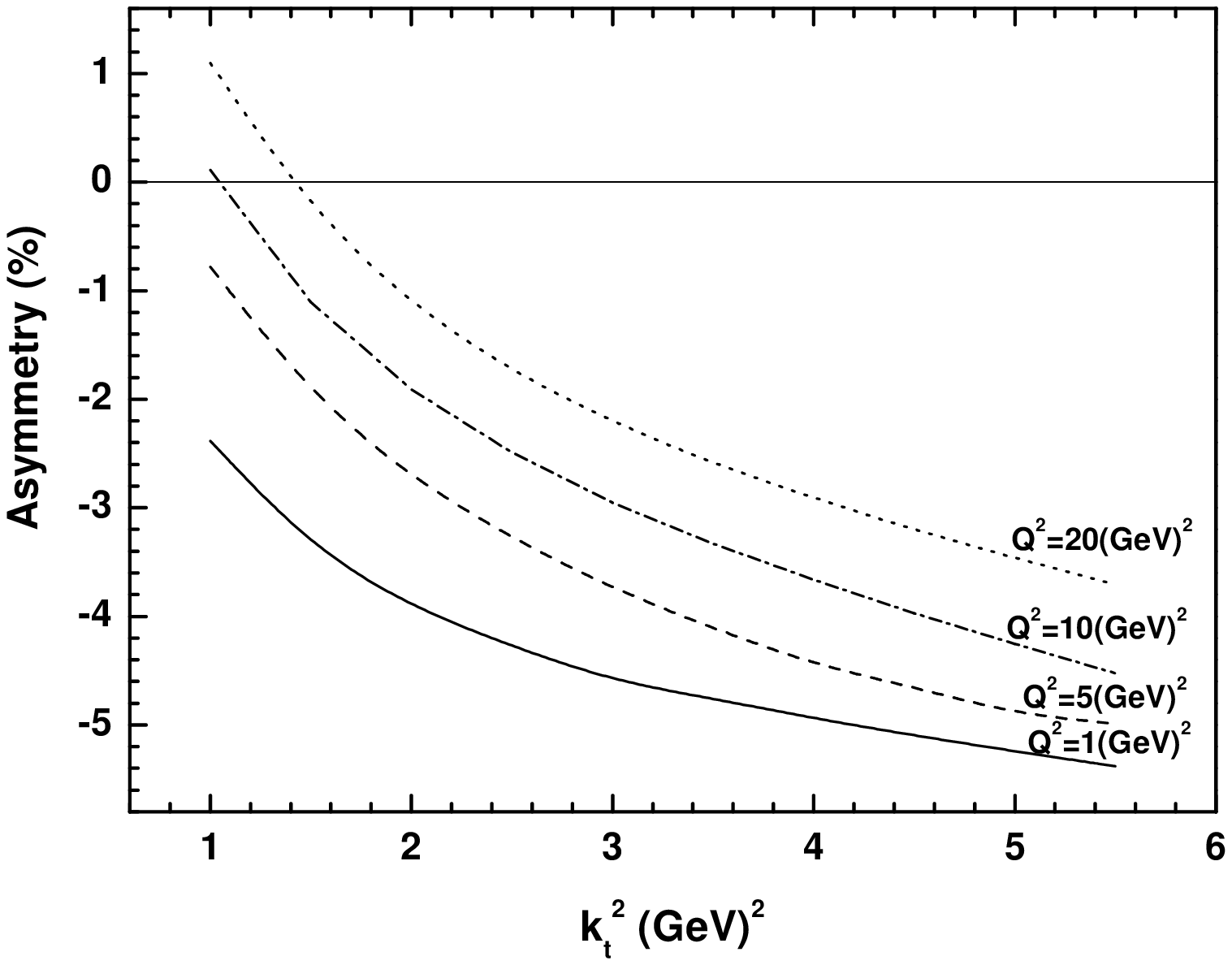}}
\caption{The $A^Q_{lT}$ asymmetry in diffractive heavy $Q\bar Q$
production at $\sqrt{s}=50 \mbox{GeV}$ for $x_P=0.05$, $y=0.3$,
$|t|=0.3 \mbox{GeV}^2$} \label{qt_h}
\end{figure}

The contribution to $A^Q_{lT}$ asymmetry which is proportional to
$ \vec Q \vec S_\perp$ is analyzed for the case when the
transverse jet momentum $\vec Q_\perp$ is parallel to the target
polarization $\vec S_\perp$ (a maximal contribution to the
asymmetry). The predicted $A^Q_{lT}$ asymmetry  in diffractive
light $Q \bar Q$ production at $\sqrt{s} =50 \mbox{GeV}$ is shown
in Fig. \ref{qt_l}. The corresponding results for heavy quarks is
presented in Fig. \ref{qt_h}. The $A^Q_{lT}$ asymmetry has a
visible mass dependence.  For light quark production, asymmetry is
not small. For $Q^2 \sim 1 \mbox{GeV}^2$ it changes the sign at
$k_\perp^2 \sim 3.5 \mbox{GeV}^2$. For heavy quark production the
predicted asymmetry is negative and not small, too.


\begin{figure}
\centering \mbox{\epsfysize=80mm\epsffile{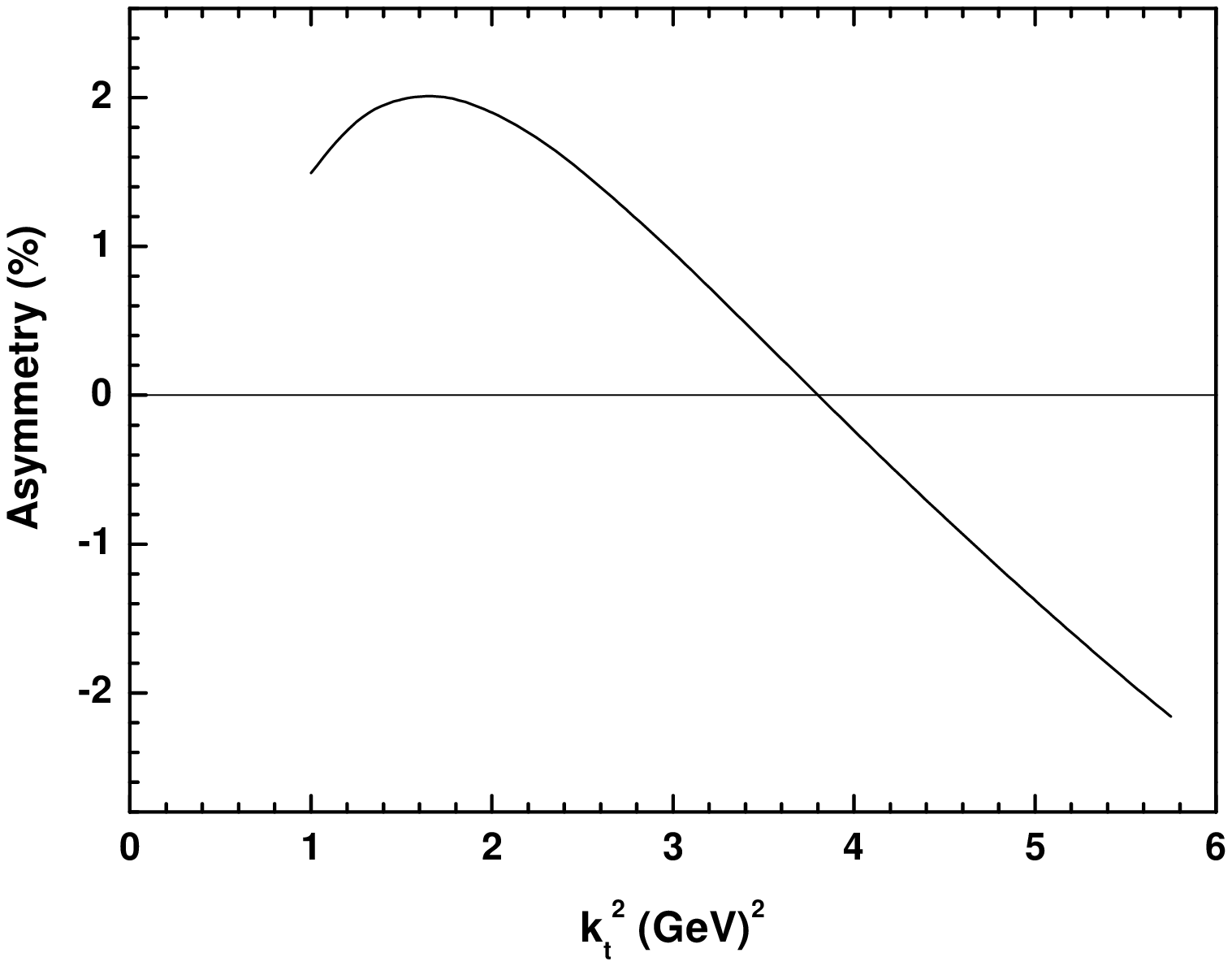}}
\caption{The $A^Q_{lT}$ asymmetry in diffractive light $Q \bar Q$
production at $\sqrt{s}=50 \mbox{GeV}$ for $x_P=0.05$, $y=0.3$,
$Q^2=5  \mbox{GeV}^2$,
 integrated over momentum transfer}
\label{iqt_l}
\end{figure}

Note that in most experiments it is difficult to detect the final
hadron (see section 2) and correspondingly to determine the hadron
momentum transfer. In this case, it is useful to have predictions
for the asymmetry integrated over momentum transfer
\begin{equation}\label{intasy}
\bar A^Q_{lT}= \frac{\int_{t_{min}}^{t_{max}}\,\sigma(-)\,dt}
{\int_{t_{min}}^{t_{max}}\,\sigma(+)\,dt}.
\end{equation}
We integrate cross sections from $t_{min} \sim 0$ up to $t_{max}=4
\mbox{GeV}^2$. The predicted integrated asymmetry for light quarks
is shown in Fig \ref{iqt_l}. It is not small, about 1-2\% for
$k_{\perp}^2=2-3 \mbox{GeV}^2$. As the nonintegrated asymmetry,
the integrated one changes the sign near $k_\perp^2 \sim 3.5
\mbox{GeV}^2$.

\section{Conclusions}

The diffractive $Q \bar Q$ leptoproduction at small $x$  which is
predominant  by the gluon exchange was analysed at eRHIC energies
within the two-gluon spin-dependent exchange model. The spin
asymmetry is found to be proportional to the ratio of polarized
gluon GPD
\begin{equation}
 A_{LT} \sim C \frac{{\cal K}^g_{x_P}(x_P)}
 {{\cal F}^g_{x_P}(x_P)}.
 \label{asylt1}
\end{equation}

The information about ${\cal K}^g$ can be obtained from asymmetry
 if the coefficient $C$ in (\ref{asylt1}) is not small.
The $A_{lT}$ asymmetry of diffractive $Q \bar Q$ production
contains two  terms which are proportional to $\vec k_\perp \vec
S_\perp$ and  to   $\vec Q \vec S_\perp$ (\ref{nm}). These terms
in the asymmetry have different kinematic properties and can be
studied independently. The term $\propto \vec k_\perp \vec
S_\perp$ has a large coefficient $C_k^{Q \bar Q}$ that is
predicted to be about 0.3-0.5. Our results for asymmetry of light
and heavy quark production in this case is quite similar. We can
conclude that this term in the asymmetry might be an excellent
tool to study transverse effects in the proton-- gluon coupling.
However, the experimental study of this asymmetry is not so
simple. To find nonzero asymmetry in this case, it is necessary to
distinguish quark and antiquark jets and to have a possibility to
analyse the azimuthal event structure over the transverse jet
momentum. The expected $A_{lT}$ asymmetry for the term $\propto
\vec Q \vec S_\perp$  is not small, too. The predicted coefficient
$C_Q^{Q \bar Q}$ in this case is about 0.1-0.2. This asymmetry is
expected to be quite different for light and heavy quark
production. The asymmetry found in the model is  about 2--4\% for
$k_\perp^2 \sim 2  \mbox{GeV}^2$ and $Q^2 \sim 1 \mbox{GeV}^2$ and
has a different sign for light and heavy quark production.

Thus, our estimations show that the corresponding coefficient $C$
in different terms of the $A_{lT}$ asymmetry might not be small.
It should be possibile to study $A_{lT}$ asymmetry in a future
eRHIC experiment for a transversely polarized proton where the
important information on the polarized gluon GPD can be
obtained.\\

The author expresses his gratitude to A.Deshpande for stimulated
discussion and remarks. I would like to thank  A. Efremov, P.
Kroll, O. Nachtmann, and O. Teryaev for fruitful discussions. This
work was supported in part by the Russian Foundation for Basic
Research, Grant 03-02-16816.



\begin{thebibliography}{9}
\bibitem{zeus97} ZEUS Collab., J. Breitweg et al, Z. Phys.
C {\bf 75}, 215 (1997).
\bibitem{jpsi1} H1 Collaboration,  S. Aid et al, Nucl. Phys.
B {\bf 472}, 3 (1996).
\bibitem{h1_99} H1 Collab., C. Adloff et al, Eur. Phys. J. C {\bf 10},  373 (1999).
\bibitem{dijet} ZEUS Collab., J. Breitweg et al, Eur. Phys. J. C {\bf 5}, 41 (1998);\\
               H1 Collab., C. Adloff et al, Eur. Phys. J. C {\bf 6},  421 (1999).
\bibitem{gol_mod}  S.V. Goloskokov, S.P. Kuleshov, O.V. Selyugin,
           Z. Phys. C {\bf 50},  455 (1991).
\bibitem{gol_kr} S.V.\ Goloskokov, P.\ Kroll, Phys.\ Rev.
          D {\bf 60}, 014019 (1999).
\bibitem{butt} N.H. Buttimore  et al,  Phys.Rev. D {\bf 59}, 114010 (1999);
\bibitem{gol_glu} S.V. Goloskokov, Phys. Lett. B {\bf 315}, 459 (1993).
\bibitem{kopel}  B.Z. Kopeliovich,  in {\it Proceedings of Spin 2002 15 th
International Spin Physics Symposium}, Brookhaven, 2002, edited by
Y.I. Makdisi, A.U. Luccio, W.W.MacKay, AIP Conf. Proc. 675, p. 58.
\bibitem{gol02} S.V.\ Goloskokov, in {\it Proceedings of Spin 2002 15 th
International Spin Physics Symposium}, Brookhaven, 2002, edited by
Y.I. Makdisi, A.U. Luccio, W.W.MacKay, AIP Conf. Proc. 675, p.
464.
\bibitem{golostr} S.V.\ Goloskokov, Euro.\ Phys.\ J. C {\bf 24},  413  (2002).
\bibitem{low} F.E.\ Low, Phys.\ Rev.\ D {\bf 12}, 163  (1975) \\
              S.\ Nussinov, Phys.\ Rev.\ Lett.\ {\bf 34}, 1286 (1975).
\bibitem{rad-j} A.V. Radyushkin, Phys.Rev. D {\bf 56}, 5524 (1997);
 X. Ji, Phys.Rev. {\bf D55} 7114 (1997).
\bibitem{die95} M.Diehl, Z. Phys. C {\bf 66}, 181 (1995).
\bibitem{bart96} J. Bartels, C. Ewerz, H. Lotter, M.W\"usthoff,
                Phys. Lett. B {\bf 386}, 389 (1996).
\bibitem{ryskin97}  E.M. Levin, A.D. Martin, M.G. Ryskin, T. Teubner,
            Z. Phys.  C {\bf 74}, 671 (1997).
\bibitem{schaef} B. Lehmann-Dronke, M. Maul, S. Schaefer, E.Stein, A. Sch\"afer,
       Phys.Lett. B {\bf 457},  207 (1999).
\bibitem{abhay} A. Deshpande, Nucl. Phys. Proc. Suppl.  {\bf 105}, 178 (2002).
\bibitem{prag03} S.V. Goloskokov, to appear in {\it Proc of the Conference
"Symmetries and Spin" - Praha-SPIN-2003}, Prague, 2003,
hep-ph/0311340.
\bibitem{krish} G. Fidecaro  et al, Phys. Lett.  B {\bf 76},
369 (1978);  B {\bf 105}, 309 (1981).
\bibitem{akch}  N. Akchurin, S.V. Goloskokov, O.V. Selyugin,
Int.J.Mod.Phys. A{\bf 14}, 253 (1999).
\bibitem{dis03} S.V.\ Goloskokov, P.Kroll, B.Postler, to appear in
{\it Proc of DIS03 workshop}, St. Petersburg,  April 2003,
hep-ph/0308140.
\bibitem{nach} T. Arens, M. Diehl, O. Nachtmann, P.V. Landshoff,
     Z. Phys. {\bf C74}, 651 (1997).
\bibitem{gehr} T.~Gehrmann and W.~J.~Stirling,
Phys.\ Rev.\ D {\bf 53}, 6100 (1996).
\end{thebibliography}
\end{document}